\documentclass[12pt,prd,aps,amssymb,amsmath,tightenlines,showpacs]{revtex4}
\usepackage{epsf,amsfonts,amsmath}
\usepackage{graphicx,float,subfigure,epsfig,color,rotating,latexsym}
\usepackage{slashed}
\usepackage[latin1]{inputenc}
\usepackage[T1]{fontenc}
\def\tr{{\rm tr}} 
\def\cP{\mathcal P}
\def\cC{\mathcal C}
\def\cT{\mathcal T}
\def\cL{\mathcal L}
\def\half{\frac{1}{2}}

\graphicspath{{Figu/}}

\begin{document}

\title{\bf Ordinary versus $\cP\cT$-symmetric $\phi^3$ quantum field theory}

\author{Carl M. Bender$^a$}\email{cmb@wustl.edu}
\author{V. Branchina$^b$}\email{branchina@ct.infn.it}
\author{Emanuele Messina$^b$}\email{emanuele.messina@ct.infn.it}

\affiliation{${}^a$Department of Physics, Kings College London, Strand, London
WC2R 1LS, UK \footnote{Permanent address: Department of Physics, Washington
University, St. Louis, MO 63130, USA.} \\
${}^b$Department of Physics, University of Catania and INFN, Sezione di Catania,
Via Santa Sofia 64, I-95123 Catania, Italy}

\date{\today}

\begin{abstract}
A quantum-mechanical theory is $\cP\cT$-symmetric if it is described by a
Hamiltonian that commutes with $\cP\cT$, where the operator $\cP$ performs space
reflection and the operator $\cT$ performs time reversal. A $\cP\cT$-symmetric
Hamiltonian often has a parametric region of unbroken $\cP\cT$ symmetry in which
the energy eigenvalues are all real. There may also be a region of broken $\cP
\cT$ symmetry in which some of the eigenvalues are complex. These regions are
separated by a phase transition that has been repeatedly observed in laboratory
experiments. This paper focuses on the properties of a $\cP\cT$-symmetric $ig
\phi^3$ quantum field theory. This quantum field theory is the analog of the
$\cP\cT$-symmetric quantum-mechanical theory described by the Hamiltonian $H=p^2
+ix^3$, whose eigenvalues have been rigorously shown to be all real. This paper
compares the renormalization-group properties of a conventional Hermitian $g
\phi^3$ quantum field theory with those of the $\cP\cT$-symmetric $ig\phi^3$
quantum field theory. It is shown that while the conventional $g\phi^3$ theory
in $d=6$ dimensions is asymptotically free, the $ig\phi^3$ theory is like a
$g\phi^4$ theory in $d=4$ dimensions; it is energetically stable, perturbatively
renormalizable, and trivial.
\end{abstract}


\maketitle

\section{Introduction}
\label{s1}

A $\cP\cT$-symmetric quantum theory is described by a Hamiltonian that commutes
with $\cP\cT$, where the operators $\cP$ and $\cT$ perform space reflection
and time reversal \cite{R1,R2}. Even if a $\cP\cT$-symmetric Hamiltonian is not
Dirac Hermitian (that is, it is not invariant under combined matrix
transposition and complex conjugation), the eigenvalues of the Hamiltonian can
still be entirely real. $\cP\cT$-symmetric Hamiltonians are particularly
interesting because they often have a parametric region of {\it unbroken} $\cP
\cT$ symmetry in which the eigenvalues are all real and a region of {\it broken}
$\cP\cT$ symmetry in which some of the eigenvalues are complex
\cite{R1,R2,R3,R4}. These regions are separated by a phase transition that has
been repeatedly observed in laboratory experiments
\cite{R5,R6,R7,R8,R9,R10,R11,R12,R13,R14}.

A heavily studied class of $\cP\cT$-symmetric Hamiltonians is \cite{R1,R2,R3,R4}
\begin{equation}
H=p^2+x^2(ix)^\epsilon,
\label{e1}
\end{equation}
where $\epsilon$ is a real parameter. The eigenvalues of this Hamiltonian are
all real when $\epsilon\geq0$ and mostly complex when $-1<\epsilon<0$. Thus, the
region of unbroken $\cP\cT$ symmetry is $\epsilon\geq0$ and the region of broken
$\cP\cT$ symmetry is $-1<\epsilon<0$. These two regions are separated by a phase
transition at $\epsilon=0$ \cite{R1,R2,R3,R4}.

A special example of a $\cP\cT$-symmetric Hamiltonian whose eigenvalues are all
real and positive is the cubic Hamiltonian
\begin{equation}
H=p^2+ix^3.
\label{e2}
\end{equation}
The $d$-dimensional, Euclidean-space, field-theoretic equivalent of this
quantum-mechanical theory is described by the Lagrangian density
\begin{equation}
\cL=\half(\partial\phi)^2+\half m^2\phi^2+i\frac{g}{6}\phi^3.
\label{e3}
\end{equation}
This Lagrangian is clearly not Hermitian, but if we assume that the field $\phi$
transforms as a pseudoscalar, then it is $\cP\cT$-symmetric. This is because
under this assumption, $\phi$ changes sign under space reflection $\cP$, and
since $i$ changes sign under $\cT$, the interaction term is $\cP\cT$ invariant.

While a conventional $g\phi^3$ theory is interesting from a theoretical point of
view, it is, of course, a physically unacceptable theory because the real cubic
potential $\half m^2\phi^2+\frac{1}{6}g\phi^3$ is not bounded below. As a
consequence, there cannot be a stable ground state.

Perturbation theory provides an easy intuitive explanation for the absence of a
stable ground state. The Feynman graphical rules for a conventional $g\phi^3$
quantum field theory follow directly from the Lagrangian density
\begin{equation}
\cL=\half(\partial\phi)^2+\half m^2\phi^2+\frac{g}{6}\phi^3.
\label{e4}
\end{equation}
The momentum-space amplitudes for a vertex and a line are
\begin{eqnarray}
{\rm vertex:}&\quad& -g,\nonumber\\
{\rm line:} &\quad& \frac{1}{p^2+m^2}.
\label{e5}
\end{eqnarray}
Using these Feynman rules, we can in principle calculate the ground-state energy
density $E_0(g)$ by summing all connected vacuum graphs. Because all such graphs
have even numbers of vertices, this sum takes the form of a formal Taylor series
in powers of $g^2$:
\begin{equation}
E_0(g)=\sum_{n=0}^\infty A_{2n}g^{2n},
\label{e6}
\end{equation}
where $A_{2n}$ is the contribution of graphs having $2n$ vertices. The key point
here is that all graphs contributing to the ground-state energy density have the
same sign and add in phase, and thus the coefficients in the series (\ref{e6})
all have the same sign. This series is divergent because the number of graphs
having $2n$ vertices grows like $n!$ \cite{R15,R16}, but unlike the perturbation
series for a $g\phi^4$ field theory, it is not a Stieltjes series \cite{R17}
because it does not alternate in sign. Consequently, the Borel sum \cite{R17} of
the perturbation series has a cut on the real-positive axis in the complex-$g^2$
plane. This perturbative argument shows that the ground-state energy density is
complex; the imaginary part of the energy density is the discontinuity across
the cut. We conclude that the ground-state of the conventional $g\phi^3$ theory
is unstable; that is, it decays (tunnels out to infinity through the barrier in
the potential) with a lifetime given by the imaginary part of $E_0(g^2)$.

On the other hand, perturbation theory also gives a simple intuitive argument
that the non-Hermitian, $\cP\cT$-symmetric Lagrangian density (\ref{e3}) defines
a theory with a stable ground state. Note that the cubic potential in this
theory is complex, and thus we cannot ask whether it is unbounded below. The
idea of a potential being bounded below applies only if the potential is real;
unlike the real numbers, the complex numbers are not ordered, so the notion of
boundedness simply does not apply. We obtain the $\cP\cT$-symmetric Lagrangian
in (\ref{e3}) from the conventional Lagrangian in (\ref{e4}) by replacing $g$ by
$ig$. When we do so, the perturbation series in (\ref{e6}) now {\it alternates
in sign}. As a consequence, it is a series of Stieltjes and its Borel sum is
{\it real} \cite{R18,R19,R20}. We conclude from this argument that it is likely
that the ground-state for this theory is stable.

While this perturbative argument is only heuristic, there is a rigorous proof
\cite{R21,R22} that the spectrum of the cubic, quantum-mechanical $\cP
\cT$-symmetric Hamiltonian in (\ref{e2}) is real and bounded below. It is not
yet known at a rigorous level whether the energy levels of the unconventional
quantum field theory in (\ref{e3}) are real and bounded below because for this
theory one can only rely on perturbative calculations.

To show that the $\cP\cT$-symmetric quantum field theory in (\ref{e3}) is a
physically acceptable quantum theory one must (in addition to proving that the
spectrum of the theory is bounded below) verify that there is a Hilbert space
with a positive inner product and that time evolution is unitary. To demonstrate
this, one would have to show that there exists a linear operator $\cC$ whose
square is unity and that $\cC$ commutes with both the Hamiltonian and with the
$\cP\cT$ operator \cite{R1,R2}. In perturbation theory the $\cC$ operator for
the $ig\phi^3$ theory has been calculated to leading order \cite{R23}, but it is
not known rigorously whether the Lagrangian (\ref{e3}) defines a physically
acceptable theory. (There may even be a critical value of $g$ at which a $\cP
\cT$ phase transition from a physically acceptable theory having real energies
to an unphysical theory having complex eigenvalues occurs.) However, we do know
for certain that the conventional $g\phi^3$ Lagrangian in (\ref{e4}) defines a
physically {\it unacceptable} theory!

While the conventional Lagrangian in (\ref{e4}) is physically unacceptable and
the unconventional Lagrangian in (\ref{e3}) may or may not be physically
acceptable, it is certainly interesting to study these Lagrangians from a
mathematical point of view. The purpose of this article is to examine and
contrast the renormalization-group properties of these two Lagrangians. We will
show that while a conventional $g\phi^3$ theory in $d=6$ dimensions is
asymptotically free, the $ig\phi^3$ theory is like a $g\phi^4$ theory in $d=4$
dimensions; that is, it is stable, perturbatively renormalizable, and trivial.

This paper is organized as follows: In Sec.~\ref{s2} we review the standard
perturbative renormalization treatment of a conventional $g\phi^3$ theory. Then,
in Sec.~\ref{s3} we carry out the renormalization-group analysis for the $g
\phi^3$ theory. In Sec.~\ref{s4} we repeat the analysis of Sec.~\ref{s3} for a
$\cP\cT$-symmetric $ig\phi^3$ theory. We give some concluding remarks in
Sec.~\ref{s5}.

\section{Perturbation theory for a $d$-dimensional $g\phi^3$ theory}
\label{s2}

The vacuum persistence functional in the presence of an external source $J$ for
a $d$-dimensional Euclidean-space quantum field theory described by a Lagrangian
$\cL$ is
\begin{equation}
Z[J]=\int\mathcal{D}\phi\,e^{\int d^dx(-\cL+J\phi)}.
\label{e7}
\end{equation}
Let us consider the unrenormalized Lagrangian for a conventional Hermitian $g
\phi^3$ quantum field theory in which we include a linear self-interaction term:
\begin{equation}
\cL=\half(\partial_\mu\phi)^2+\half m^2\phi^2+\frac{g}{6}\phi^3+h\phi.
\label{e8}
\end{equation}
We can then rewrite $Z[J]$ as
\begin{equation}
Z[J]=\mathcal{N}e^{-\int V(\delta/\delta J)}e^{\frac{1}{2}\int\int J D_b J},
\label{e9}
\end{equation}
where $\mathcal{N}$ is a normalization constant, $D_b$ is the usual bosonic
propagator in coordinate space, and $V(\phi)=h\phi+g\phi^3/6$.

The one-loop one-particle-irreducible unrenormalized vertex functions in
momentum space are
\begin{eqnarray}
\Gamma^{(1)}&=&h+\frac{g}{2}\int\frac{d^d p}{(2\pi)^d}\frac{1}{p^2+m^2},
\label{e10}\\
\Gamma^{(2)}(q)&=& q^2+m^2-\frac{g^2}{2}\int\frac{d^d p}{(2\pi)^d}\frac{1}{(
p^2+m^2)[(p+q)^2+m^2]},
\label{e11}\\
\Gamma^{(3)}(q_1,q_2)&=& g+g^3\int\frac{d^d p}{(2\pi)^d}\frac{1}{(p^2+m^2)
[(p+q_1)^2+m^2][(p+q_1+q_2)^2+m^2]}.
\label{e12}
\end{eqnarray}
To evaluate the above integrals we use the standard integral identities
\begin{eqnarray}
\int\frac{d^dl}{(2\pi)^d}\frac{1}{(l^2+\Delta)^n}&=&\frac{1}{(4\pi)^{d/2}}
\frac{\Gamma(n-d/2)}{\Gamma(n)}\Delta^{\frac{d}{2}-n},
\label{e13}\\
\int\frac{d^dl}{(2\pi)^d}\frac{l^2}{(l^2+\Delta)^n}&=&
\frac{1}{(4\pi)^{d/2}}\frac{d}{2}\frac{\Gamma(n-d/2-1)}{\Gamma(n)}\Delta^{
\frac{d}{2}+1-n},
\label{e14}\\
\int \frac{d^dl}{(2\pi)^d}\frac{l^{\mu}l^{\nu}}{(l^2+\Delta)^n}
&=&\int\frac{d^dl}{(2\pi)^d}\frac{l^2\eta^{\mu\nu}/\tr(\eta)}{(l^2+\Delta)^n},
\label{e15}\\
\int\frac{d^dl}{(2\pi)^d}\frac{l^{\mu}}{(l^2+\Delta)^n}&=&0.
\label{e16}
\end{eqnarray}
where $\eta^{\mu\nu}$ is the metric matrix.

The upper critical dimension for the Hermitian $g\phi^3$ theory is $d=6$. At $d=
6$ the cubic operator $\phi^3$ is marginal (just as $\phi^4$ is marginal at $d=
4$). The theory turns out to be asymptotically free, as we will see below.

Normally, in textbooks the $\phi^3$ theory at or near $d=6$ dimensions is
discussed for pedagogical reasons \cite{R24}. This is because the perturbative
results are easily established and the theory provides a simple example of an
asymptotically free theory. Furthermore, unlike the $g\phi^4$ theory in $d=4$
dimensions, a contribution to the wave function renormalization constant $Z$ is
already present at the one-loop level. However, no physical meaning is attached
to the conventional $g\phi^3$ theory because, as noted earlier, it is unstable
(that is, the spectrum is unbounded below).

Let us now examine the behavior of this $g\phi^3$ theory near $d=6$. Let $I_1$,
$I_2$, and $I_3$ represent the three integrals that appear in $\Gamma_1$,
$\Gamma_2$, and $\Gamma_3$ above. With the help of (\ref{e13}), at $d=6-
\epsilon$ we get
\begin{equation}
I_1=\int\frac{d^d p}{(2\pi)^d}\frac{1}{p^2+m^2}=\frac{m^4\mu^{-\epsilon}}{64
\pi^3\epsilon}+{\rm O}\left(\epsilon^0\right),
\label{e17}
\end{equation}
where here and in the following we introduce the 't Hooft scale $\mu$ and give
only the divergent parts of $I_1$, $I_2$ and $I_3$.

Next, we consider the second integral
\begin{equation}
I_2=\int\frac{d^d p}{(2\pi)^d}\frac{1}{(p^2+m^2)[(p+q)^2+m^2]}.
\label{e18}
\end{equation}
To extract its divergent part, we take two derivatives:
\begin{eqnarray}
\frac{\partial I_2}{\partial q^{\mu}}&=&-\int\frac{d^d p}{(2\pi)^d}\frac{2(p+q
)_{\mu}}{(p^2+m^2)[(p+q)^2+m^2]^2},\nonumber\\
\frac{\partial^2 I_2}{\partial q^{\mu}\partial q^{\nu}}&=& 
\int\frac{d^d p}{(2\pi)^d}\frac{8(p+q)_{\mu}(p+q)_{\nu}-2g_{\mu\nu}[(p+q)^2
+m^2]}{(p^2+m^2)[(p+q)^2+m^2]^3}.
\label{e19}
\end{eqnarray}
We then expand $I_2(q)$ around $q=0$:
\begin{eqnarray}
I_2(q)&=&I_2\Big|_{q=0}+q^{\mu}\left.\frac{\partial I_2}{\partial q^\mu}
\right|_{q=0}+\frac{1}{2}q^\mu q^\nu\left.\frac{\partial^2 I_2}{\partial q^\mu
\partial q^\nu}\right|_{q=0}+I_2(q)^{\rm (finite)}\nonumber\\
&=&\int\frac{d^d p}{(2\pi)^d}\frac{1}{(p^2+m^2)^2}-\int\frac{d^d p}{(2\pi)^d}
\frac{2q\cdot p}{(p^2+m^2)^3}\nonumber\\
&&\quad+\int\frac{d^d p}{(2\pi)^d}\frac{4(p\cdot q)^2-q^2(p^2+m^2)}
{(p^2+m^2)^4}+I_2(q)^{\rm (finite)}\nonumber\\
&=&\int \frac{d^d p}{(2\pi)^d}\frac{1}{(p^2+m^2)^2}
-\int\frac{d^d p}{(2\pi)^d}\frac{1}{(p^2+m^2)^3}\nonumber\\
&&\quad+\frac{4}{\tr\eta}q^2\int
\frac{d^d p}{(2\pi)^d}\frac{p^2}{(p^2+m^2)^4}+I_2(q)^{\rm (finite)},
\label{e20}
\end{eqnarray}
where we have used the identities (\ref{e13}) and (\ref{e14}). The result is
\begin{eqnarray}
I_2(q)&=&-\frac{q^2\mu^{-\epsilon}}{192\pi^3\epsilon}-\frac{m^2\mu^{-\epsilon}}
{32\pi^3 \epsilon}+{\rm O}\left(\epsilon^0\right).
\label{e21}
\end{eqnarray}

Finally, for $I_3$ we use the identity
\begin{eqnarray}
I_3&=&\int \frac{d^d p}{(2\pi)^d}\frac{1}{(p^2+m^2)[(p+q_1)^2+m^2][(p+q_1+q_2)^2
]+m^2}\nonumber\\
&=&\int_0^1 dx\,dy\,dz\,\delta(x+y+z-1)\,\int\frac{d^d p}{(2\pi)^d}\frac{1}{D^3}
\label{e22}
\end{eqnarray}
in which $D$ is evaluated at $k=q_1+q_2$:
\begin{eqnarray}
D&=&x(p^2+m^2)+y[(p+q_1)^2+m^2]+z[(p+k)^2+m^2]\nonumber\\
&=&(x+y+z)(p^2+m^2)+2 p\cdot(y q_1+z k)+y q_1^2+z k^2.
\label{e23}
\end{eqnarray}
By performing the shift $l=p+y q_1+z k$, $D$ becomes
\begin{equation}
D=l^2+m^2+yq_1^2+z k^2-(yq_1+z k)^2.
\label{e24}
\end{equation}
We then obtain
\begin{eqnarray}
I_3&=&\int_0^1 dz\int^{1-z}_0 dy\int\frac{d^d l}{(2\pi)^d}\frac{2}{\left[l^2+
m^2+yq_1^2+zk^2-(yq_1+zk)^2\right]^3}\nonumber\\
&=&\frac{\mu^{-\epsilon}}{64\pi^3\epsilon}+{\rm O}\left(\epsilon^0\right).
\label{e25}
\end{eqnarray}

In terms of the standard definitions for the renormalized quantities
\begin{eqnarray}
\phi&=&Z^{1/2}\phi_R,\nonumber\\
Z&=&1+\delta Z,\nonumber\\
h&=&Z^{-1/2}(h_R+\delta h),\nonumber\\
m^2&=&Z^{-1}(m_R^2 + \delta m^2),\nonumber\\
g&=&Z^{-3/2}(\mu^{\epsilon/2}g_R+\delta g),
\label{e26}
\end{eqnarray}
where the $\phi^3$ coupling constant $g_R$ is made dimensionless by introducing
the 't Hooft scale $\mu$, the renormalized vertex functions are
\begin{eqnarray}
\Gamma_R^{(1)}&=&h_R+\delta h-\frac{g_Rm^4_R\mu^{-\epsilon/2}}{128\pi^3\epsilon}
+\dots,\nonumber\\
\Gamma_R^{(2)}&=&p^2+m^2_R+\delta Zp^2+\delta m^2+g_R^2\left(\frac{p^2}{384\pi^3
\epsilon}+\frac{m_R^2}{64\pi^3\epsilon}\right)+\dots,\nonumber\\
\Gamma_R^{(3)}&=&g_R+\delta g+\frac{g_R^3\mu^{\epsilon/2}}{64\pi^3\epsilon}
+\dots,
\label{e27}
\end{eqnarray}
where we have omitted the finite one-loop contributions. Therefore, by adopting
the $MS$-scheme \cite{R25}, we get
\begin{eqnarray}
\delta h&=&\frac{g_Rm^4_R\mu^{-\epsilon/2}}{128\pi^3\epsilon},\nonumber\\
\delta Z&=&-\frac{g_R^2}{384\pi^3\epsilon},\nonumber\\
\delta m^2&=&-\frac{g_R^2 m_R^2}{64\pi^3\epsilon},\nonumber\\
\delta g&=&-\frac{g_R^3 \mu^{\epsilon/2}}{64\pi^3\epsilon}.
\label{e28}
\end{eqnarray}

Finally, we define the {\it dimensionless} renormalized couplings $h$, $m^2$,
and $g$, which should not be confused with the bare parameters in (\ref{e26}):
\begin{eqnarray}
h_R&=&\mu^{4-\epsilon/2}h,\nonumber\\
m^2_R&=&\mu^{2}m^2,\nonumber\\
g_R&=&g.
\label{e29}
\end{eqnarray}
The one-loop renormalization-group (RG) functions for the dimensionless
renormalized couplings are then given by
\begin{eqnarray}
\gamma&=&\half\mu\frac{\partial}{\partial\mu}\delta Z=\frac{g^2}{768\pi^3},
\label{e30}\\
\beta_h&=&-(4-\epsilon/2)h-\mu h\frac{\partial(\mu^{\epsilon/2-4}\delta h
/h)}{\partial \mu}+\gamma h\nonumber\\
&=&-(4-\epsilon/2)h+\frac{g m^4}{128\pi^3}+\frac{g^2 h}{768\pi^3},
\label{e31}\\ 
\beta_{m^2}&=&-2m^2-\mu m^2\frac{\partial(\mu^{-2}\delta m^2/m^2)}{\partial\mu}
+2\gamma m^2\nonumber\\
&=&-2m^2-\frac{g^2m^2}{64\pi^3}+\frac{g^2m^2}{384\pi^3}=-2m^2-\frac{5g^2m^2}
{384\pi^3},
\label{e32}\\
\beta_g&=&-\frac{\epsilon}{2}g-\mu g\frac{\partial(\mu^{-\epsilon/2}\delta g/g)}
{\partial\mu}+3\gamma g\nonumber\\
&=&-\frac{\epsilon}{2}g-\frac{g^3}{64\pi^3}+\frac{g^3}{256\pi^3}=-\frac{
\epsilon}{2}g-\frac{3 g^3}{256\pi^3}.
\label{e33}
\end{eqnarray}

\section{Renormalization-group analysis of $g\phi^3$ theory}
\label{s3}

From (\ref{e31}), (\ref{e32}), and (\ref{e33}), we see that near $d=6$ the
theory possesses only a Gaussian fixed point (GFP): $h^*={m^2}^*=g^*=0$. As is
well known, the linearization of the RG equations around the GFP shows that
near this point the couplings scale according to their scaling dimension. That
is, by defining $t=\ln(\mu/\mu_0)$, we find that
\begin{eqnarray}
h(t)&\sim&e^{-(4-\epsilon/2)t},\nonumber\\
m^2(t)&\sim&e^{-2 t},\nonumber\\
g(t)&\sim& e^{-\epsilon t/2}.
\label{e34}
\end{eqnarray}
Note that $\gamma(g^*)=0$ at the GFP.

Finally, from $\beta_g$ we see that at $d=6$ the theory is asymptotically free.
The explicit solution of the RG equation $\mu\frac{d(g^2)}{d\mu}=2g\beta_g$ is
\begin{equation}
g^2(\mu)=\frac{g_0^2}{1+\frac{3g_0^2}{128\pi^3}\ln\left(\frac{\mu}{\mu_0}
\right)},
\label{e35}
\end{equation}
where $\mu_0$ is an arbitrary scale and $g_0=g(\mu_0)$. In (\ref{e35}) we
immediately recognize the usual features of asymptotic freedom and infrared
slavery. However, we emphasize that despite exhibiting these important physical
properties, the conventional $g\phi^3$ theory is unstable. 

\section{RG analysis for the $\cP\cT$-symmetric $ig\phi^3$ theory}
\label{s4}

By making the substitutions $h\to ih$ and $g\to ig$ in (\ref{e31})--(\ref{e33}),
we find that 
\begin{eqnarray}
\gamma&=&-\frac{g^2}{768\pi^3},
\label{e36}\\
\beta_h&=&-(4-\epsilon/2)h+\frac{gm^4}{128\pi^3}-\frac{g^2h}{768\pi^3},
\label{e37}\\ 
\beta_{m^2}&=&-2m^2+\frac{5g^2m^2}{384\pi^3},
\label{e38}\\
\beta_g&=&-\frac{\epsilon}{2}g+\frac{3g^3}{256\pi^3}.
\label{e39}
\end{eqnarray}
Unlike the conventional $g\phi^3$ theory, we now have {\it nontrivial fixed
points} at
\begin{eqnarray}
h^{*}&=&0,\nonumber\\
m^{2*}&=&0,\nonumber\\
g^{*}&=&\pm\sqrt{128\pi^3\epsilon/3},
\label{e40}
\end{eqnarray}
in addition to the usual GFP.

\begin{figure}[!ht]
\epsfxsize=13cm
\centerline{\epsffile{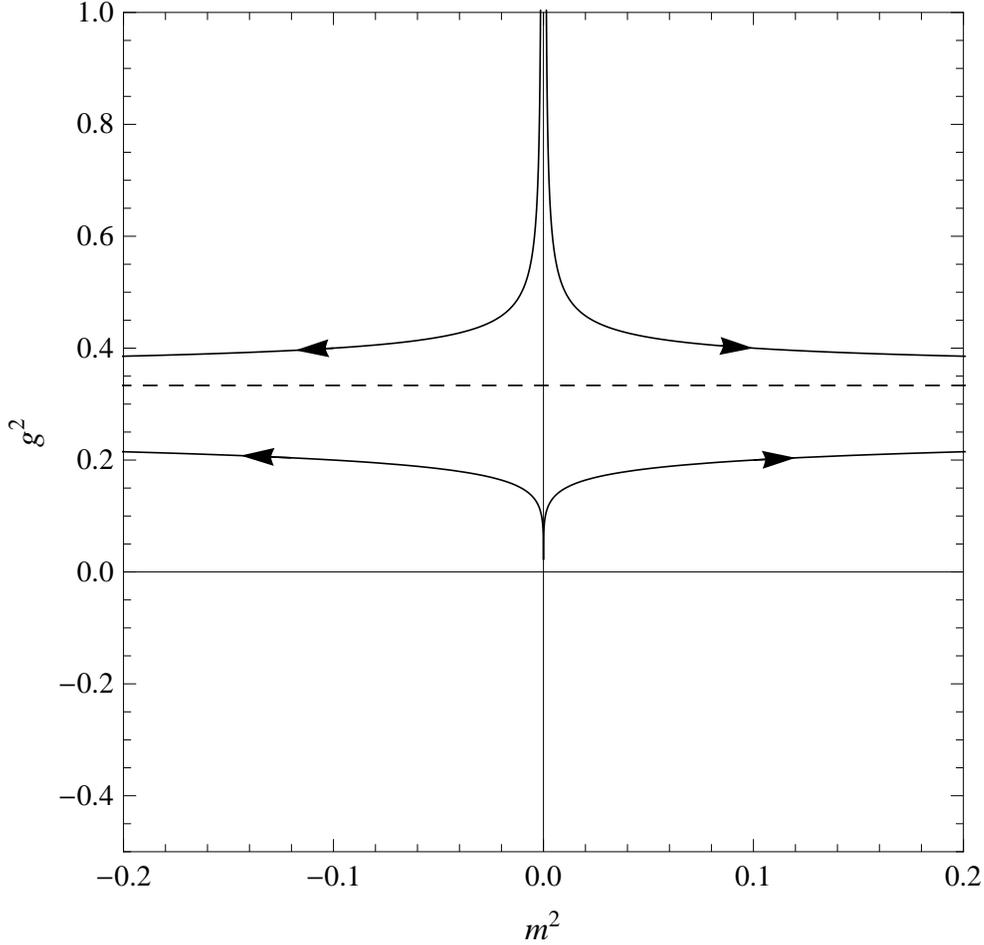}}
\caption{Four RG trajectories in the $(m^2,g^2)$ plane near the non-Gaussian 
fixed point $m^{2*}=0$, $g^{2*}=128\pi^3\epsilon/3$ obtained from (\ref{e38})
and (\ref{e39}) for $\epsilon=0.5$. The four initial values are $m^2(t=0)=-0.1,
\,0.1,\,-0.1,\,0.1$ and correspondingly $g^2(t=0)=0.2,\,0.4,\,0.4,\,0.2$. The
eigendirections are the dashed line and the $g^2$ axis.}
\label{F1}
\end{figure}

As in the conventional case, the flow near the GFP is dictated by the canonical
dimensions of the couplings. Near the non-Gaussian fixed points, however, the
linearization of the RG equations gives the following {\it new scaling
behavior}:
\begin{eqnarray}
h(t)&=&c_1e^{g_1 t},\nonumber\\
m^2(t)&=&c_2 e^{g_2 t},\nonumber\\
g(t)&=&g^*+c_3 e^{g_3 t},
\label{e41}
\end{eqnarray}
where $g_1=(-4+4\epsilon/9)$, $g_2=(-2+5\epsilon/9)$ and $g_3=\epsilon$ are the
eigenvalues of the $3\times 3$ Jacobian matrix that defines the linearized RG
flow around the non-Gaussian fixed points, and $c_1$, $c_2$, and $c_3$ are
arbitrary coefficients. This result comes from solving the linearized system of
differential RG equations around the non-Gaussian fixed points (see
Fig.~\ref{F1}). From these equations we see that $h$, $m^2$, and $g$ are still
eigendirections of the Jacobian matrix, as was the case for the GFP. Finally,
the anomalous dimension of the field is
\begin{eqnarray}
\gamma=-\frac{\epsilon}{18}.
\label{e42}
\end{eqnarray}

It is worth noting that the hyperscaling relation that connects the anomalous
dimension of the field with the eigenvalue $g_1$, namely
\begin{eqnarray}
\eta=2\gamma=2+d+2g_1,
\label{e43}
\end{eqnarray}
is satisfied, as expected. Here, $\eta$ is the exponent that gives the anomalous
scaling of the two-point function. Near the critical region, the latter behaves
as
\begin{eqnarray}
\Gamma_R^{(2)}(q)\sim\frac{1}{q^{2-\eta}}.
\label{e44}
\end{eqnarray}

\section{Conclusions} 
\label{s5}

We have shown that the $\cP\cT$-symmetric $ig\phi^3$ quantum field theory near
$d=6$ dimensions possesses three fixed points, the GFP and two nontrivial ones
in (\ref{e40}). At $d=6$ ($\epsilon=0$) the three fixed points merge in a unique
fixed point, which is the gaussian one. From the $\beta_g$ function (\ref{e39}),
we can see that when $\epsilon=0$, the theory is trivial:
\begin{eqnarray}
g^2(\mu)&=&\frac{g_0^2}{1-\frac{3 g_0^2}{128\pi^3}\ln\left(\frac{\mu}{\mu_0}
\right)}.
\label{e45}
\end{eqnarray}
This allows us to conclude that the $ig\phi^3$ theory is energetically {\it
stable}, {\it perturbatively renormalizable}, and {\it trivial}. This triviality
property is the same as for the conventional Hermitian $g\phi^4$ theory in $d=4$
dimensions. If we consider this $ig\phi^3$ theory in $d=6$ dimensions from an
effective-field-theory standpoint (as is the case for the Higgs sector of the
Standard Model), it can be treated as a perfectly sensible physical theory. 

\begin{figure}[!ht]
\epsfxsize=13cm
\centerline{\epsffile{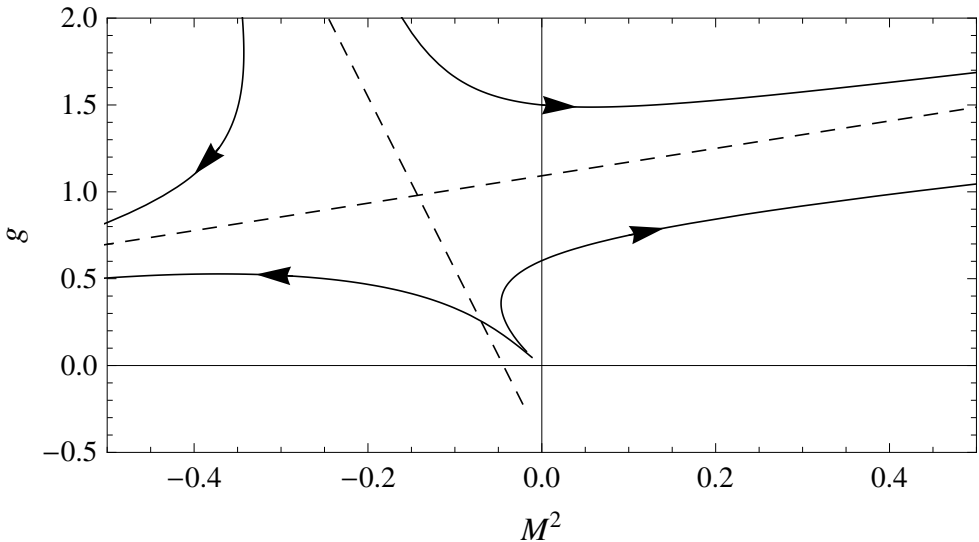}}
\caption{Four RG trajectories in the $(M^2,\,g)$ plane for the scalar $g\phi^4$
theory in $d=3$ dimensions near the Wilson-Fisher fixed point. The initial
values are: $M^2(t=0)=-0.25,\,0.1,\,-0.4,\,0$ and correspondingly $g=0.5,\,0.75,
\,1.1,\,1.5$. The eigendirections are indicated by the two dashed lines.}
\label{F2}
\end{figure}

From the RG point of view, however, what seems to us to be more interesting is
what happens when $d<6$ ($d=6-\epsilon$). In this case, if we consider the
$(m^2,g^2)$ plane, we have a situation that closely parallels the ferromagnetic
case as described in $d=4-\epsilon$ dimensions, where we have the {\it Gaussian}
and the {\it Wilson-Fisher} fixed points. In Fig.~\ref{F2} the ($M^2,\,g$) plane
for the ordinary $g\phi^4$ theory in $d=4-\epsilon$ dimensions is shown and the
RG flows on this plane are plotted. The GFP is at the origin, while the
Wilson-Fisher fixed point is on the left of the $M^2=0$ axis. The dashed lines
are the eigendirections and the Wilson-Fisher fixed point is at the crossing of
the two eigendirections (one of which is the relevant direction, the other the
irrelevant one). The two fixed points, the GFP and the WFFP determine the RG
flows on this plane. In the case of a $\cP\cT$-symmetric $ig\phi^3$ theory in $d
=6-\epsilon$ dimensions the situation in the $(m^2,\,g^2)$ plane is essentially
the same. However, the role of the $M^2$ term of the ferromagnetic model is
played by $m^2$, while the role of $g$ (in the $g\phi^4$ term) is played by
$g^2$ (compare Figs.~\ref{F1} and \ref{F2}). In the $(m^2,\,g^2)$ plane the two
eigendirections are the $m^2=0$ axis and the dashed line of Fig.~\ref{F1}. The
non-Gaussian fixed point is at the crossing of the two eigendirections. 

It is evident from Figs.~\ref{F1} and \ref{F2} that the RG flow in the ($m^2$,
$g^2$) plane is the {\it same} as the RG flow in the ferromagnetic case; that
is, it is the same as the flow in the ($M^2,\,g$) plane. In both cases these
flows are governed by the two fixed points (the Gaussian one and the
non-Gaussian one). As is clear from Figs.~\ref{F1} and \ref{F2}, the Gaussian
fixed point of the ferromagnetic case corresponds to the Gaussian fixed point of
the $ig\phi^3$ theory; the Wilson-Fisher fixed point of the ferromagnetic case
corresponds to our non-Gaussian fixed point: $m^2=0$, $g^2=128\pi^3\epsilon/3$.
Regarding the two fixed points in (\ref{e40}), $g=\pm\sqrt{128\pi^3\epsilon/3}$,
it should be noted that in establishing this parallel, we refer to the square of
the coupling constant $g^2$ rather than to the coupling constant $g$ itself. We
do this because it is convenient to treat the two fixed points in a unified
manner because the physics around either fixed point is the same.

We note that the additional non-Gaussian fixed points of the $\cP\cT$-symmetric
theory are also present in the conventional $g\phi^3$ theory although they are
purely imaginary [see $\beta_g$ in (\ref{e33})]. Therefore, by considering also
the purely imaginary solutions to the equation $\beta_g =0$ in the conventional
$g\phi^3$ theory, in a sense we recover the results obtained by stating from the
beginning that the $g\phi^3$ coupling in the Lagrangian is purely imaginary
(which is the case for the $\cP\cT$-symmetric theory). In summary, while the
equation $\beta_g=0$ in the conventional $g\phi^3$ theory has one real and two
imaginary conjugate solutions, in the $\cP\cT$-symmetric theory all of the three
solutions are real.
 
Finally, we point out that in both the conventional and the $\cP\cT$-symmetric
theories the RG equations for $g$ and $m^2$ with $\beta_{m^2}$ and $\beta_g$
given by (\ref{e32})--(\ref{e33}) and (\ref{e38})--(\ref{e39}), respectively,
can be solved exactly. Having defined $t=\ln\frac{\mu}{\mu_0}$ as before, for
the conventional theory we get
\begin{eqnarray}
g^2(t)&=& \frac{g_0^2 e^{-\epsilon t}}{1+\frac{3 g_0^2}{128\pi^3}\left(\frac{1-
e^{-\epsilon t}}{\epsilon}\right)},
\label{e46}\\
m^2(t)&=& m_0^2 e^{-2t}\left[1+\frac{3 g_0^2}{128\pi^3}\left(\frac{1-e^{-
\epsilon t}}{\epsilon}\right)\right]^{-5/9},
\label{e47}
\end{eqnarray}
and for the $\cP\cT$-symmetric theory we get
\begin{eqnarray}
g^2(t)&=&\frac{g_0^2 e^{-\epsilon t}}{1-\frac{3 g_0^2}{128\pi^3}\left(\frac{1-
e^{-\epsilon t}}{\epsilon}\right)},
\label{e48}\\
m^2(t)&=& m_0^2 e^{-2t}\left[1-\frac{3 g_0^2}{128\pi^3}\left(\frac{1-e^{-
\epsilon t}}{\epsilon}\right)\right]^{-5/9}.
\label{e49}
\end{eqnarray}

\acknowledgments

CMB thanks the U.S.~Department of Energy and the Leverhulme Foundation and EM
thanks the Centro Siciliano di Fisica Nucleare e Struttura della Materia
(CSFNSM) for financial support.


\begin{thebibliography}{99}

\bibitem{R1} C.~M.~Bender, {\it Contemp.~Phys.} {\bf 46}, 277 (2005).

\bibitem{R2} C.~M.~Bender {\it Repts.~Prog.~Phys.}, {\bf 70}, 947 (2007).

\bibitem{R3} C.~M.~Bender and Boettcher S, {\it Phys.~Rev.~Lett.} {\bf 80},
5243 (1998).

\bibitem{R4} C.~M.~Bender, S.~Boettcher, and P.~N.~Meisinger, {\it J. Math.
Phys.} {\bf 40}, 2201 (1999).

\bibitem{R5} A.~Guo, G.~J.~Salamo, D.~Duchesne, R.~Morandotti,
M.~Volatier-Ravat, V.~Aimez, G.~A.~Siviloglou, and D.~N.~Christodoulides, {\it
Phys. Rev. Lett.} {\bf 103} 093902 (2009).

\bibitem{R6} C.~E.~R\"uter, K.~G.~Makris, R.~El-Ganainy, D.~N.~Christodoulides,
M.~Segev, and D.~Kip, {\it Nat. Phys.} {\bf 6}, 192 (2010).

\bibitem{R7} J.~Rubinstein, P.~Sternberg, and Q.~Ma, {\it Phys. Rev. Lett.} {\bf
99}, 167003 (2007).

\bibitem{R8} K.~F.~Zhao, M.~Schaden, and Z.~Wu, {\it Phys. Rev.} A {\bf 81},
042903 (2010).

\bibitem{R9} Y.~D.~Chong, L.~Ge, and A.~D.~Stone {\it Phys. Rev. Lett.} {\bf
106}, 093902 (2011).

\bibitem{R10} Z.~Lin, H.~Ramezani, T.~Eichelkraut, T.~Kottos, H.~Cao, and
D.~N.~Christodoulides {\it Phys. Rev. Lett.} {\bf 106}, 213901 (2011).

\bibitem{R11} C.~Zheng, L.~Hao, and G.~L.~Long, arXiv:1105.6157 [quant-ph].

\bibitem{R12} S.~Bittner, B.~Dietz, U.~Guenther, H.~L.~Harney, M.~Miski-Oglu,
A.~Richter, and F.~Schaefer, Phys.~Rev.~Lett. (in press, 2012).

\bibitem{R13} J.~Schindler, A.~Li, M.~C.~Zheng, F.~M.~Ellis, T.~Kottos, {\it
Phys. Rev.} A {\bf 84}, 040101 (2011).

\bibitem{R14} Optical PT lattices: A.~Szameit, M.~C.~Rechtsman,
O.~Bahat-Treidel, and M.~Segev, Phys.~Rev.~A {\bf 84}, 021806(R) (2011).

\bibitem{R15} C.~M.~Bender and T.~T.~Wu {\it Phys.~Rev.~Lett.}~{\bf 27}, 461
(1971).

\bibitem{R16}  C.~M.~Bender and T.~T.~Wu, {\it Phys.~Rev.} D {\bf 7}, 1620
(1973); Reprinted in J.~C.~Le Guillou and J.~Zinn-Justin (eds.), {\it
Large-order behaviour of perturbation theory} (North-Holland, Amsterdam, 1990),
p.~41-57.

\bibitem{R17} C.~M.~Bender and S.~A.~Orszag, {\it Advanced Mathematical Methods
for Scientists and Engineers} (McGraw Hill, New York, 1978).

\bibitem{R18} E.~Caliceti, S.~Graffi, and M.~Maioli, Comm.~Math.~Phys.~{\bf 75},
51 (1980).

\bibitem{R19} C.~M.~Bender and G.~V.~Dunne, J.~Math.~Phys.~{\bf 40}, 4616
(1999).

\bibitem{R20} C.~M.~Bender and E.~J.~Weniger, J.~Math.~Phys.~{\bf 42}, 2167
(2001).

\bibitem{R21} P.~Dorey, C.~Dunning,and R.~Tateo, {\it J.~Phys.~A:
Math.~Gen.} {\bf 34}, L391 and {\bf 34}, 5679 (2001).

\bibitem{R22} P.~Dorey, C.~Dunning, and R.~Tateo, {\it J.~Phys.~A:
Math.~Gen.} {\bf 40}, R205 (2007).

\bibitem{R23} C.~M.~Bender, D.~C.~Brody, and H.~F.~Jones, Phys.~Rev.~Lett.~{\bf
93}, 251601 (2004).

\bibitem{R24}  See for instance M.~Peskin and D.~Schroeder, {\it An Introduction
To Quantum Field Theory} (Addison-Wesley, New York, 1995).

\bibitem{R25} See for instance J.~Collins, {\it Renormalization: An Introduction
to Renormalization} (Cambridge University, Cambridge, 1984) and J.~Zinn-Justin,
{\it Quantum Field Theory and Critical Phenomena} (Claredon, New York, 1996).

\end{thebibliography}
\end{document}